\documentclass[prd,nofootinbib,twocolumn,showpacs]{revtex4}
\usepackage{graphics}
\usepackage{bm}

\begin{document}

\title{
Symmetries within domain walls
}

\author{Tanmay Vachaspati}
\affiliation{Department of Physics, Case Western Reserve University,
10900 Euclid Avenue, Cleveland, OH 44106-7079, USA.}

\begin{abstract}
The comparison of symmetries in the interior and the exterior 
of a domain wall is relevant when discussing the correspondence
between domain walls and branes, and also when studying the 
interaction of walls and magnetic monopoles. I discuss the symmetries
in the context of an $SU(N)\times Z_2$ model (for odd $N$) with a 
single adjoint scalar field. Situations in which the wall interior 
has less symmetry than the vacuum are easy to construct while the 
reverse situation requires significant engineering of the scalar 
potential. 
\end{abstract}


\

\maketitle

\section{Introduction}

A connection between fundamental branes and field theoretic domain 
walls was drawn in \cite{DvaShi97,DvaVil02}. Processes occurring
in brane scenarios were shown to have direct analogs in 
topological defect scenarios and vice versa (though often
with effects such as confinement taken into account). This 
connection is noteworthy because defects have been studied for 
several decades and are relatively well-understood systems. 
One property that is central to brane scenarios is that matter
as we know it is confined to the brane. Only gravitons can 
propagate in the bulk. The analog of this property in the defect 
setting is that there are massless excitations on a domain wall
and that these excitations cannot escape the wall because they
are massive outside. The precise spectrum of massless excitations
will depend on the symmetries inside and outside the domain wall.

The spectrum of excitations that live on a topological defect is 
also very important for determining the interaction of the defect
with other defects in the system. As an example, consider the
``sweeping scenario'' described in \cite{DvaLiuVac97} where
domain walls collide with magnetic monopoles that are produced
in the same phase transition. The outcome of the collision
depends on whether the magnetic monopole can unwind inside 
the domain wall \cite{PogVac00a,PogVac00b}. This in turn depends
on the symmetry that exists inside the wall and whether it is
large enough for the monopole to unwind. If the symmetry is
sufficiently large, the walls can sweep away all the monopoles
and potentially solve the cosmological magnetic monopole 
overabundance problem. The interactions of walls and point 
defects is also relevant in the laboratory where both types 
of defects are observed.

In this paper, I will investigate the symmetries within 
domain walls in a model based on a 
$G_N \equiv SU(N)\times Z_2$ ($N$=odd)
symmetry group with a single adjoint scalar field, $\Phi$. 
(Symmetries within domain walls in other models have been 
investigated in \cite{AxePer97}, while other defects have
been considered in \cite{AxePerTom98,AxePerTro98,CarBraDav02}.)
The 
Lagrangian of our model is:
\begin{equation}
L = {\rm Tr} (\partial_\mu \Phi )^2 - V(\Phi ) 
\label{lagrangian}
\end{equation}
where $V(\Phi )$ is a potential invariant under $G_N$. The
$Z_2$ factor in $G_N$ takes $\Phi \rightarrow - \Phi$. 
If the vacuum expectation value (VEV) of $\Phi$ is:
\begin{equation}
\Phi_0 = \eta \sqrt{2 \over {N(N^2-1)}}
                  \pmatrix{n{\bf 1}_{n+1}&{\bf 0}\cr
                      {\bf 0}&-(n+1){\bf 1}_n\cr} \ ,
\label{phi0}
\end{equation}
where ${\bf 1}_p$ is the $p\times p$ identity matrix and $\eta$ is
an energy scale determined by the minima of the scalar 
potential $V(\Phi )$, the symmetry breaking pattern is:
\begin{equation}
H_N = [SU(n+1)\times SU(n)\times U(1)]/C \ .
\label{unbrokensymm} 
\end{equation}
The group $C$ is the center of $H$ and $N \equiv 2n+1$.

The symmetry breaking $G_N \rightarrow H_N$ leads to domain walls
since the discrete $Z_2$ factor in $G_N$ is spontaneously broken.
Therefore, if on one side of the wall if $\Phi = \Phi_0$, then
on the other side we can have $\Phi =  - U^\dag \Phi_0 U$ for 
any choice of $U \in SU(N)$. The freedom in choosing $U$ leads
to $n+1$ different topological domain wall solutions in the model 
\cite{PogVac01}, each of which has different symmetry restoration
within. More specifically, let
\begin{equation}
\Phi_- \equiv \Phi (x = -\infty ) = \Phi_0
\label{Phiatminusinfty}
\end{equation}
Then, we can choose
\begin{eqnarray}
&{}&\Phi_+ \equiv \Phi (x=+\infty ) =
 - \eta \sqrt{2 \over {N(N^2-1)}} \times \nonumber \\
       &{}& {\rm diag} ( n{\bf 1}_{n+1-q}, -(n+1){\bf 1}_{q},
                         n{\bf 1}_{q}, -(n+1){\bf 1}_{n-q}) 
\label{phi+choices}
\end{eqnarray}
where $q=0,...,n$ tells us how many diagonal entries of $\Phi_-$ 
have been permuted in $\Phi_+$. The case $q=0$ is when 
$\Phi_+ = - \Phi_-$ and corresponds to the $Z_2$ wall in a
$\phi^4$ model embedded in this more complicated model. 

The solution for any $q$ can be written as:
\begin{equation}
\Phi (x) = F_+ (x) {\bf M_+} + F_- (x) {\bf M_-} + g(x) {\bf M}\ ,
\label{kinkexplicit2}
\end{equation}
where
\begin{equation}
{\bf M}_+ =  {{\Phi_+ + \Phi_-}\over {2}} \ , \ \ {\bf M}_- =
{{\Phi_+ - \Phi_-}\over {2}} \label{M+M-} \ .
\end{equation}
The matrix $M$ is found by requiring that the solution
extremize the energy functional. This can be done for an
arbitrary polynomial potential $V(\Phi )$ that is invariant
under $G_N$. This gives: 
\begin{eqnarray}
{\bf M} = &\mu& \, {\rm diag} ( q(n-q){\bf 1}_{n+1-q}, \nonumber \\
   &-&(n-q)(n+1-q){\bf 1}_{2q} , q(n+1-q){\bf 1}_{n-q} )
\label{Mresult}
\end{eqnarray}
with $\mu$ being a normalization factor in which we also include
the energy scale $\eta$ for convenience:
\begin{equation}
\mu = \eta [ 2q(n-q)(n+1-q)\{ 2n(n+1-q)-q\} ]^{-1/2} \ .
\label{muvalue}
\end{equation}
The expression for ${\bf M}$ does not hold for $q=0$ and 
$q=n$. In these cases, we have $g(x) =0$ and ${\bf M}$ drops
out of the solution. 

It will be convenient to also explicitly list the values
of ${\bf M}_\pm$:
\begin{eqnarray}
{\bf M}_+ = \eta &N& \sqrt{1 \over {2N(N^2-1)}} \nonumber \\
&{}& {\rm diag} ( 0_{n+1-q}, {\bf 1}_q, -{\bf 1}_q, 0_{n-q} )
\label{M+} \ ,
\end{eqnarray}
\begin{eqnarray}
{\bf M}_- = &\eta&  \sqrt{1 \over {2N(N^2-1)}} \nonumber \\
&{}&{\rm diag} ( -2n {\bf 1}_{n+1-q}, {\bf 1}_{2q},
                 2(n+1){\bf 1}_{n-q} ) \ .
\label{M-}
\end{eqnarray}

The boundary conditions on the profile functions are:
$F_+ (\pm \infty ) = + 1$, $F_- (\pm \infty )= \pm 1$
and $g(\pm \infty ) = 0$. To determine the interior symmetry
of the wall, we are interested in finding the values of the
profile functions at the center of the wall which we take
to be $x=0$.

The antisymmetry of $F_-(x)$, namely $F_- (-x) = - F_- (+x)$,
follows from the symmetry of the equations of motion (for
general potential $V(\Phi )$) and the boundary conditions.
This immediately tells us that $F_- (x=0) = 0$. It is also
clear that $F_+ (x)$ need not vanish on the wall. To determine
$g(x=0)$ requires more care since $g(\pm \infty) =0$.
However, from the explicit expressions for the matrices
${\bf M}_-$ and ${\bf M}$ we see that:
\begin{equation}
{\rm Tr}( {\bf M_-}^{2p+1} {\bf M} ) \ne 0
\label{M+M}
\end{equation}
for any integer $p \ge 1$. Hence if $V(\Phi )$ contains 
any term of the form $\Phi^{2p+2}$, then the equation of
motion for $g(x)$ will necessarily contain a term of the
kind $[F_-(x)]^{2p+1}$. This is odd under $x \rightarrow -x$
and hence $g(x)$ must also satisfy $g(-x) = -g(x)$. This
tells us that $g(x=0) =0$. Putting all this together, we
have:
\begin{equation}
\Phi (x=0) = F_+ (0) {\bf M}_+
\label{Phiatx=0}
\end{equation}
It is important to emphasize that this result holds for 
any choice of polynomial $V(\Phi )$ that gives the desired 
symmetry breaking.

Once we know the value of $\Phi$ inside domain wall, it is
straightforward to determine the symmetry since we only need
to find those generators of $SU(N)$ that commute with
$\Phi (0)$. From the explicit expression for ${\bf M}_+$,
we see that there is a square block of zero entries of
size $2n+1-2q = N-2q$, another block proportional to
${\bf 1}_q$ and another to $-{\bf 1}_q$. Therefore
the symmetry within the wall is:
\begin{equation}
G_{in}^q = \{ SU(N-2q)\times [SU(q)\times U(1)]^2 \} /C' \ , 
\label{Ginside}
\end{equation}
where $q = 1,...,n-1$ and
$C'$ denotes the center. The two $U(1)$ factors
have been inserted since transformations given by all 
diagonal generators remain unbroken by $\Phi (x=0)$.

The above result for $G_{in}^q$ holds when $q\ne 0$.
This case can be treated individually.
If $q =0$, we see that $F_+ (x) =0$, and so we have:
\begin{equation}
G_{in}^0 = SU(N) \ .
\label{Ginside0}
\end{equation}

Next we determine the values of $q$ for which the interior
symmetry exceeds the exterior (bulk) symmetry. The number
of generators of the exterior symmetry group, $z_{out}$, 
are determined from eq. (\ref{phi+choices}):
\begin{equation}
z_{out} = [(n+1)^2-1]+[n^2-1]+1={1\over 2}(N^2-1)
\label{zout}
\end{equation}
The number of generators of the interior symmetry group,
$z_{in}$, are found from $G_{in}^q$:
\begin{eqnarray}
z_{in} &=& [(N-2q)^2-1]+ 2[q^2-1+1] \nonumber \\
       &=& 6q^2 -4qN + (N^2-1) \ , \ \ q=0,...,n 
\label{zin}
\end{eqnarray}
Therefore the difference in the number of generators of
the interior and exterior symmetries are:
\begin{equation}
\Delta z^q \equiv z_{in}-z_{out} 
           = 6q^2-4qN + {1\over 2}(N^2-1) \ , 
\label{Deltazq}
\end{equation}
for $q = 0,...,(N-1)/2$.
The expression for $\Delta z^q$ is quadratic in $q$ and we can 
easily find that to get a larger symmetry inside the wall than 
outside ($\Delta z^q > 0$, $q \le (N-1)/2$), we need
\begin{equation}
q < {{2N - \sqrt{N^2+3}}\over 6}
\label{qbound}
\end{equation}
{}For large values of $N$, we find $q \lesssim N/6$. As an example,
if $N=5$, then only the $q=0$ wall has symmetry restoration in
its interior; all the other walls have symmetry reduction in
their interior.

We now ask if the $q=0$ wall can be stable for any $N$ for some
choice of potential $V(\Phi )$. We can address this question in
two ways: first we can check the perturbative stability of the
$q=0$ wall, and second, we can compare the mass of the $q=0$
wall to the $q=n$ wall. If the $q=0$ wall is perturbatively
stable but more massive than the $q=n$ wall, then we would
expect that in a realistic setting, a network of $q=0$ walls
will eventually decay into a network of $q=n$ walls. So both
local and global stability are of interest.

If the potential $V(\Phi )$ is quartic, the analysis in 
\cite{PogVac00b,PogVac01,Vac01} shows that the $q=n$ wall
is stable while the $q=0$ wall is unstable. Here we would like
to consider the more difficult problem of a general potential.
We will derive certain conditions that $V(\Phi )$ must satisfy
if the $q=0$ wall is to be stable and/or lighter than the
$q=n$ wall.

{}First let us consider perturbative stability of the $q=0$
wall. We perturb the wall solution
\begin{equation}
\Phi (x) = f(x) T_0 + \psi (x) T
\label{perturbedkink}
\end{equation}
where $T_0 \equiv \Phi_0/\eta$, $f(x)$ denotes the wall profile 
function, $\psi (x)$ is the perturbation profile, and $T$ is a 
generator of $SU(N)$ that is chosen to be orthogonal to $\Phi_0$:
\begin{equation}
{\rm Tr} (T_0 T) = 0
\label{trT0t}
\end{equation}
Then, we find that the perturbation causes a change in
the energy
\begin{equation}
\delta E [\psi ] = {{1} \over 2} \int_{-\infty}^{+\infty} dx ~
  \psi \biggl [ - {{d^2}\over {dx^2}} + V_T^{(2)} (f) \biggr ] \psi
\label{deltaEpsi}
\end{equation}
where $V_T^{(2)} (f)$ denotes the second order variation of 
$V(\Phi )$ in $\psi$. In general, we need to determine if the 
Schrodinger equation 
\begin{equation}
 \biggl [ - {{d^2}\over {dx^2}} + V_T^{(2)} (f) \biggr ] \psi = 
                  \epsilon \psi
\label{schrodinger}
\end{equation}
has any negative eigenvalues (bound states). This will depend on
the details of the potential $V$ as well as the profile function
$f$.

In the specific case of a quartic potential, the Schrodinger 
equation has been solved \cite{PogVac00b,PogVac01} 
and bound states have been found to be present. The analysis
there suggests that it might be possible to derive a simpler
criterion for the occurrence of bound states. The idea is that
if the perturbation is along the $\Phi_0$ direction, that is,
if we choose $T_0 = T$, then the corresponding Schrodinger equation
necessarily has a zero mode because we know that translations 
of the wall do not change the energy. Furthermore the zero
mode eigenfunction is proportional to the derivative of $f(x)$. 
Therefore:
\begin{equation}
 \biggl [ - {{d^2}\over {dx^2}} + V_0^{(2)} (f) \biggr ] 
           {{df}\over {dx}} = 0
\label{schrodinger0}
\end{equation}
where $V_0^{(2)}$ denotes the second order variation of
$V$ along the $\Phi_0$ direction. A useful condition for
checking for an instability is obtained by evaluating 
$\delta E$ in eq. (\ref{deltaEpsi}) when $\psi$ is the translational 
zero mode. This leads to:
\begin{equation}
\delta E [ f'(x) ] = 
 {1\over 2} \int_{-\infty}^{+\infty} dx ~ f'(x) 
                   [ V_T^{(2)} (f) - V_0^{(2)} (f) ] f'(x)
\label{deltaEfprime}
\end{equation}
where a prime denotes derivative with respect to $x$. The
integrand still depends on the profile function and, with
another trick, we can eliminate this dependence.

The trick is to write the Bogomolnyi equation for the $q=0$
wall \cite{Bog76}. This is derived
by writing the energy functional for the $q=0$ wall in the 
following way:
\begin{equation}
E[f] = {1\over 2}\int dx ~ \biggl [ (f' - \sqrt{2V(f)} ~ )^2
                         + 2\sqrt{2V(f)} f' \biggr ]
\label{towardsbogo}
\end{equation}
The last term is a boundary term and its integration yields
the energy of the $q=0$ wall:
\begin{equation}
E = \sqrt{2} \int_{-1}^{+1} df ~ \sqrt{V(f)}
\label{bogoenergy}
\end{equation}
Minimization of the whole square term in the energy functional
yields the Bogomolnyi equation:
\begin{equation}
f'(x) = \sqrt{2 V_0 (f)}
\label{bogoeq}
\end{equation}
where $V_0 (f)$ is the potential evaluated with $\Phi = f(x)T_0$.
Note that $f'(x)$ is always positive since we are assuming that
the $V(\Phi ) = 0$ is a global minimum that occurs only when
$\Phi = \Phi_0$ (up to symmetry transformations). Therefore $f(x)$
is monotonic and we can transform the integration variable in
eq. (\ref{deltaEfprime}) to $f$:
\begin{equation}
\delta E [ f'(x) ] = 
 {1\over {\sqrt{2}}} \int_{-\eta}^{+\eta} df ~
                   [ V_T^{(2)} (f) - V_0^{(2)} (f) ] \sqrt{V_0 (f)}
\label{deltaEfprimewrtf}
\end{equation}
A sufficient condition for the instability of the $q=0$ wall is:
\begin{equation}
 \int_{-\eta}^{+\eta} df ~ 
        [ V_T^{(2)} (f) - V_0^{(2)} (f) ] \sqrt{V_0 (f)} ~ < 0
\label{instabilitycondition}
\end{equation}

A desirable feature of this instability condition is that it depends 
only on the potential $V(\Phi )$ and its second order variations. To 
check for this particular instability, we need not solve any
differential equations, but only evaluate the integral.
We expect the wall to be most unstable to Goldstone modes as
these are massless outside the wall and tachyonic within the wall.
Hence we expect that $\delta E$ will be smallest when $[T,\Phi_0] \ne 0$. 

The instability condition in eq. (\ref{instabilitycondition})
has been written down for a very specific perturbation, namely
for $\psi = f'$, while a full blown stability analysis requires
that we solve the Schrodinger equation in eq. (\ref{schrodinger}),
which also depends on the exact profile function. Hence there is 
a danger that our instability condition may not be useful and
large classes of unstable walls may violate the condition. As we
now describe, this fear is not realized, at least in the case of 
quartic potential but for arbitrary $N$. So there is hope that the
condition will be useful even for more complicated potentials.

The quartic potential for arbitrary $N$ can be written as:
\begin{equation}
V(\Phi ) = - m^2 {\rm Tr}[ \Phi ^2 ]+ h ( {\rm Tr}[\Phi ^2  ])^2 +
  \lambda {\rm Tr}[\Phi ^4]  - V_0 
\label{quarticV}
\end{equation}
where, $V_0$ is a constant chosen so that $V(\Phi )=0$ at its
global minimum. The second order variation, $V^{(2)}_T(f)$, 
must be of the form:
\begin{equation}
V^{(2)}_T(f) = a_T + b_T f^2
\label{V2form}
\end{equation}
where $a_T$ and $b_T$ may be $T$-dependent coefficients.
Now since the symmetry is completely restored when
$f=0$, $a_T =a$ which is independent of $T$. (In fact,
$a = -m^2$.) Therefore 
\begin{equation}
V^{(2)}_T(f) - V^{(2)}_0 (f) =  (b_T - b_0) f^2
\label{V2-V2form}
\end{equation}
Now if we choose $T$ so that the modes along $T$ are the 
Goldstone boson modes, then they are massless when $f=\eta$
and $V^{(2)}_T(\eta ) =0$. On the other hand, the modes
along $T_0$ are massive when $f=\eta$ and so
$V^{(2)}_0 (\eta ) > 0$. This shows that we must
have $b_T - b_0 < 0$. Hence
\begin{equation}
V^{(2)}_T(f) - V^{(2)}_0 (f) < 0 \ ,
\label{V2-V2lessthan0}
\end{equation}
for all values of $f$. Therefore the integrand in 
eq. (\ref{instabilitycondition}) is negative everywhere
and the $q=0$ is perturbatively unstable for quartic
$V(\Phi )$ for any value of $N$.

To understand the instability for quartic potentials,
we note that the center of the $q=0$ wall is at a local
maximum of the (multi-dimensional) potential. So the
wall is given by a path that goes over the maximum of
the potential, and there is at least one direction
orthogonal to the path. (It helps to imagine a path
over a Mexican hat potential in two dimensions.) So
the path can slip from the top of the potential and
the energy of the solution can be lowered. 

{}For potentials that are more complicated than quartic,
we find:
\begin{equation}
V^{(2)}_T(f) - V^{(2)}_0 (f) =  
              f^2 \sum_{n=0}^\infty \alpha_n f^{2n}
\label{V2-V2generalV}
\end{equation}
where the $\alpha_n$ are coefficients that depend on
$T$ and the choice of potential. As above, for Goldstone
modes, by setting $f=\eta$, we find:
\begin{equation}
\sum_{n=0}^\infty \alpha_n \eta^{2n} ~ < ~ 0
\label{generalcondition}
\end{equation}
Other than this condition, the potential needs to have
a global minimum with the correct unbroken symmetry.
This condition is hard to implement in general. However,
it appears that by suitably tuning the parameters in
$V(\Phi )$ it should be possible to have a $q=0$ wall
that is perturbatively stable. In other words, if
the potential is more complicated than just quartic,
the area where the slipping can occur may be very
small and gradient energy might be able to prevent 
the instability.

Even if the $q=0$ wall is perturbatively stable, it may
be more massive than other walls in the model. So we now
consider global stability by comparing the energy of
the $q=0$ wall to that of the $q=n$ wall.

First we give a heuristic arguments that indicates that 
$q=n$ is in many situations likely to be the most stable 
wall. If we write:
\begin{equation}
\Phi = {\rm diag} (f_1, f_2,...,f_N)
\label{Phiinfrep}
\end{equation}
then the boundary conditions for the $q=n$ wall require the 
least change in the $f_i$ as the wall is crossed. The 
expressions for $\Phi_\pm$ (eqs. (\ref{Phiatminusinfty}) 
and (\ref{phi+choices})) show that only one of the $f_i$ 
has to change sign and thus pass through zero. The other 
components merely have to shift a little bit. On the contrary,
the boundary conditions for the $q=0$ wall require that every 
component $f_i$ change to $-f_i$ and thus pass through zero. 
In other words, the charge of the wall is given by 
$Q = \Phi_+ - \Phi_-$ and this has large components for 
the $q=0$ wall but small components for the $q=n$ wall. 
If the energy of the wall is directly related to the charge 
(for example as ${\rm Tr}(Q^2)$) then the $q=n$ wall will have 
the least energy.

In models motivated by supersymmetry, the energy of the walls 
can be evaluated explicitly and we can then determine which wall
is lightest. In such models, we have:
\begin{equation}
V = {1\over 4}{\rm Tr} \biggl | {{dW}\over {d\Phi}} \biggr |^2
\label{susysqrtV}
\end{equation}
where $W(\Phi )$ is the ``superpotential''. By a slight
generalization of the Bogomolnyi derivation described above,
we obtain the energy of a $q$-wall:
\begin{equation}
E_{wall} = 
  \int dx ~ {\rm Tr} \biggl ( 
  {{d\Phi}\over {dx}} {{dW}\over {d\Phi}} \biggr ) 
=  W(\Phi_+^{(q)}) - W (\Phi_0)  
\label{susywallenergy}
\end{equation}
where we have used eq. (\ref{Phiatminusinfty}). 
Note that the superpotential need not be invariant under 
$SU(N)\times Z_2$; only the potential $V(\Phi )$, obtained
by differentiating $W$ and then squaring, is required to be
invariant under the symmetry group. However, if we require
that $V(\Phi_0)=0$, then also $V(\Phi_U)=0$ where 
$\Phi_U \equiv U^\dag \Phi_0 U $, $U\in SU(N)$. Using
eq. (\ref{susysqrtV}), we get
\begin{equation}
{{dW}\over {d\Phi}} \bigg | _{\Phi_U}  =0 
\label{dWoverdPhi}
\end{equation}
Since $U$ is an element of a continuous group, $\Phi_0$ and 
$\Phi_U$ are continuously related, and, $W(\Phi) = W(\Phi_U)$. 
Hence $W(\Phi_+^{(q)})$ is independent of $q$. This implies
that the energy of walls in these models does not depend
on $q$. 

The stability analysis of $q=0$ walls and the heuristic
argument based on the energy of a wall being proportional
to its charge, both suggest that the stable walls in
$SU(N)\times Z_2$ models have $q=n$. The arguments using
supersymmetry show that there exist potentials for which
the $q=0$ wall may not be unstable to decay into the $q=n$
wall. It is conceivable that this is a limiting case and any 
departures from supersymmetry might result in the $q=0$ wall 
being heavier than the $q=n$ wall.

The $q=n$ walls have an internal symmetry that is smaller than 
the bulk symmetry. Yet, for the brane analogy and the monopole 
sweeping scenario, it is desirable to have large internal symmetry 
as compared to the bulk symmetry. This can in principle happen if 
the potential $V(\Phi )$ is sufficiently complicated. The explicit 
construction of potentials that can lead to stable walls with 
large internal symmetry, remains an open problem. In Ref.
\cite{DvaShi97,DvaVil02} the situation where the exterior
non-Abelian symmetries are confining was discussed. In this situation, 
it is possible that excitations are much more massive in 
the exterior than in the interior. Another possibility for having 
enhanced interior symmetry is if the bulk symmetry is broken down, 
say completely. Then the internal symmetry must necessarily be larger 
than the bulk symmetry. However it is not clear in this situation 
if the internal symmetry can include continuous symmetries or whether, 
as in the case of the $Z_2$ kink in the $\lambda \phi^4$ model, only 
the discrete $Z_2$ will be restored within the wall. 

The conclusion that the $q=n$ wall is most likely the stable
wall in the model has consequences for the monopole sweeping
scenario \cite{PogVac00b}. Now there are certain monopoles that
can unwind within the wall and others that cannot. The ones
that cannot unwind, pass through the wall. The monopoles that
can unwind will either unwind or get rotated into monopoles
that cannot unwind. The interactions of monopoles and walls
is an important topic for further investigation.

\begin{acknowledgments} 
I am grateful to Gia Dvali, Arthur Lue, Levon Pogosian and 
Alex Vilenkin for discussions.
This work was supported by DOE grant number DEFG0295ER40898 
at CWRU.

\end{acknowledgments}

\end{document}